\documentclass[a4paper,12pt]{article}
\usepackage{amssymb,amsmath,mathbbol,mathrsfs}
\usepackage[usenames,dvipsnames]{color}
\usepackage{hyperref}
\usepackage{xcolor}
\usepackage{stmaryrd}
\usepackage{authblk}
\usepackage{framed}
\usepackage{empheq} 
\usepackage{slashed}
\usepackage{hyphenat}
\usepackage{cite}
\usepackage{apacite}
\usepackage{natbib}
\usepackage{dsfont}
\usepackage{chngcntr}
\usepackage{enumerate}  
\usepackage{tikz-cd}
\usepackage{xcolor}
\usepackage{booktabs}
\usepackage{array}

\usepackage{marginnote}    

\usepackage[left=.8in,right=.8in,top=.8in,bottom=.8in]{geometry}                  

\usepackage{sectsty} 
\allsectionsfont{\sffamily\mdseries\upshape} 
\usepackage{tocloft}

\makeatletter
\renewenvironment{abstract}{%
    \if@twocolumn
      \section*{\abstractname}%
    \else 
      \begin{center}%
        {\bfseries\sffamily\abstractname\vspace{\z@}}
      \end{center}%
      \quotation
    \fi}
    {\if@twocolumn\else\endquotation\fi}
\makeatother

\numberwithin{equation}{section}
\hypersetup{
	colorlinks=true,         
	linkcolor=MidnightBlue,          
	citecolor=BrickRed,        
	urlcolor=MidnightBlue            
}

\newcommand{\be}{\begin{equation}}
\newcommand{\ee}{\end{equation}}

\renewcommand{\d}{{\mathrm{d}}}
\newcommand{\D}{{\mathrm{D}}}

\newcommand{\Ad}{{\mathrm{Ad}}}

\renewcommand{\bar}{\overline}

\renewcommand{\hat}{\widehat}

\newcommand{\RR}{\mathds{R}} 

\newcommand{\lbr}{\llbracket}
\newcommand{\rbr}{\rrbracket}
\newtheorem{defi}{Definition}

\newcommand{\cint}{{\int\kern-.87em{<}}}
\newcommand{\sint}{{\int\kern-.75em{\sim}}}
\newcommand{\fint}{{\int\kern-1.00em{\int}}}

\newcommand{\bb}{\mathbb}

\renewcommand{\#}{\sharp}

\let\oldmarginpar\marginpar
\renewcommand\marginpar[1]{\oldmarginpar{\color{red}\raggedright\footnotesize #1}}

\title{Gauge theory is about the geometry of internal spaces }
\author{Henrique Gomes\footnote{gomes.ha@gmail.com}\\
Oriel College, University of Oxford}

\begin{document}
\maketitle
\begin{abstract} In general relativity, the strong equivalence principle  is underpinned by a geometrical interpretation of fields on spacetime: all fields and bodies probe the same geometry. This geometric interpretation  implies that the parallel transport of all spacetime tensors and spinors is dictated by a single affine connection. Can something similar be said about gauge theory? Agreed, in gauge theory  different symmetry groups rule the interactions of  different types of charges, so we cannot expect to find the same kind of universality found in the gravitational case.  Nonetheless, the parallel transport of all the fields that are charged   under the same symmetry group  is dictated by a single `gauge connection\rq{}, and they all transform jointly under a gauge transformation. Is this kind of `restricted universality\rq{} as geometrically underpinned as in  general relativity? Here I argue that it is. The key difference is that the  gauge geometry concerns `internal\rq{}, as opposed to  `external\rq{}, spaces. The gauge symmetry of the standard model  is thus understood as merely the  automorphism group of an internal geometric structure---$\bb C^3\otimes \bb C^2\otimes \bb C^1$ endowed with an orientation and canonical inner product---in  the same way as  spacetime symmetries (such as Poincar\'e transformations), are understood  as the automorphism group of an external geometric structure (respectively, a Minkowski metric). And the Ehresmann connection can then be understood as determining parallelism for   this internal geometry.
\end{abstract}
\tableofcontents
\section{Introduction}\label{sec:intro}

All quarks and gluons interact via one and the same strong nuclear force. Notwithstanding this interaction, in their modern mathematical guise, these particles exist as sections of distinct vector bundles over spacetime. Cashing this out in more colloquial terms,  these particles are described by  fields that take values in a variety of internal spaces co-existing over each spacetime point. Nonetheless, the fields  interact because these internal spaces don\rq{}t just coexist over each spacetime point: they are intimately connected. For instance, we are not free to perform independent gauge transformations on  quarks and gluons; and were we to take particular values for quarks and gluons at spacetime point $p$ and bring them to  point $q$ along a certain spacetime curve $\gamma$, these values would  perform a `synchronised rotation\rq{} in their respective internal spaces, their evolution `marches in step\rq{}.  

As we will see in Section \ref{sec:parallel}, the usual mathematical explanation for this  co-rotation is that the vector bundles of which these fields are sections are \emph{associated vector bundles}; and all these bundles are associated to a single principal $G$-bundle, $P$--- where $G$ is the symmetry group regimenting a particular interaction: $SU(3)$ in the case of the strong force, $SO(3,1)$ in the case of general relativity, where each is endowed with a single Ehresmann affine connection $\omega$. 

Thus, in the standard model of particle physics (SM henceforth),  all fields charged under the same gauge group get their parallel transport from the same mathematical object, $\omega$. But  $\omega$ does not inhabit any of the spaces of the physical matter fields; it exists on a more abstract space, viz. a principal fiber bundle $P$,  and its sole function is to coordinate the parallel transport of the different physical fields in their respective internal spaces. In the words of \citet[p. 2401]{Weatherall2016_YMGR}:
\begin{quote}
Principal bundles are auxiliary [in the sense that only] vector bundles represent possible local states of matter; principal bundles coordinate between these vector bundles ... [they are auxiliary] in the sense in which a coach is auxiliary to the players on the field.
\end{quote}
This is a beguiling metaphor, but is it explanatory? It certainly falls short of the geometric explanation for symmetry and parallel transport that we get in general relativity.

 In this paper, by using parallel transport in general relativity as a template, I will provide  better explanations for the gauge case. The most apt formulation  of general relativity to be used as a template employs an orthonormal basis of vectors at each spacetime point and a connection-form that describes their parallel transport. The different orthonormal bases are related by  elements of the  Lie group $O(3,1)$ (or $SO(3,1)$, if spacetime orientation is important): the group that leaves the Minkowski metric on a $3+1$ space invariant (and its subgroup of orientation preserving transformations).  In other words, the transformation group is tied to the preservation of the structure of a `typical fiber\rq{}: which is a vector space over each spacetime point---viz. the tangent space, which is isomorphic to $\RR^{4}$---endowed with a semi-definite inner product. 
 
 Using this template,  two putative disanalogies between gravity and gauge   that seem to hinder  the gauge case having a similar  explanation become apparent.

  The first putative disanalogy is that principal bundles in the gravitational case are entirely dispensable, unlike  in the gauge case. For, in the gauge case, we need to invoke principal bundles in order to have a unified interpretation of parallel transport and gauge transformations: that all fields charged under the same gauge group get their parallel transport from the same mathematical object, $\omega$.  On the other hand,  in the case of general relativity,  in order to describe parallel transport of vector fields on the spacetime manifold $M$ we need not invoke principal bundles at all. This is right, as far as tensor fields are concerned: the putative principal bundle in question would be a (sub)-bundle of the bundle of frames of the tangent spaces, but we don\rq{}t need to invoke it, since we don\rq{}t need frames to describe parallel transport of tensor fields (recall how general relativity textbooks expound tensor analysis, by having the metric determine the notion of parallel transport (i.e. the Levi-Civita connection), with never a mention of a principal fibre bundle).

 To dispel this first disanalogy,  I recall the fact that matter fields in classical field theory involve sections of spinor bundles, not only of tensor bundles. And spinors  are most naturally articulated using frames and the frame bundle: as in gauge theory, this ensures their structure group plays an important role in their description. Moreover, spinors inherit the notion of parallel transport from the tangent bundle $TM$, but their geometric relation to $TM$ is elaborate, and it requires the use of additional structure on $TM$, like an inner product and orientation. So classical spinors illustrate the possibly elaborate vector bundles that can be geometrically based on a fundamental typical fiber that  represents (locally of course) the external space for general relativity. I will argue that similar constructions are available in the `internal\rq{} space in the gauge case.  So, in both the internal and the external cases,  there are fields that benefit from a description using frames and structure groups, but this description is not strictly speaking necessary: both cases can be understood purely geometrically without mention of frames and groups. 

  The second putative disanalogy is as follows.  In general relativity sections of different tensor and spinor bundles march in step under parallel transport \emph{because} they are all constructed from the same geometric structures: namely, the tangent bundle $TM$, with each tangent space $T_pM$ being endowed with a Lorentzian inner product and an orientation (necessary in the case of spinors). In the gauge case, the textbook tradition---indeed, so far as I know, the extant literature\footnote{I thank Lathan Boyle and David Tong for helpful discussions of this curious lacuna in the literature.}---reveals no similarly powerful explanation for why the fields that couple through the strong force march in step under parallel transport. We merely---but very successfully!---stipulate that some of the fields are associated to the same principal bundle (and therefore the same connection $\omega$). In Weatherall\rq{}s vivid metaphor, we assign a single coach to all of the players. And while it is true that the symmetry group on the bundle preserves some structure---that of the principal bundle---the object whose structure is preserved  is not itself part of the furniture of the world; it is used only to coordinate the properties of the physical fields. In sum:  it is clear why in general relativity the same Levi-Civita connection should guide the parallel transport of different tensor fields; it is clear why they march in step. Whereas it is \emph{not} clear why the same Ehresmann connection should guide the parallel transport of different gauge fields; and it is not clear what \emph{physical} structure is preserved by the gauge symmetry group.  
  
    Similarly, I will dispel the second disanalogy by showing that one \emph{can} see all the constituents of both the  chromodynamics and the electroweak sector of the SM as sections of  vector bundles whose typical fibers are constructed from the same geometric structure: namely $\mathbb{C}^3\otimes\mathbb{C}^2\otimes \mathbb{C}$,   endowed with the canonical inner product of the complex planes and an orientation. As we will see (in Section \ref{sec:admissible}), the great advantage of seeing a principal bundle $P$ as a bundle of frames of a vector bundle $E$ is that the gauge group is no longer postulated as fundamental; it acquires meaning as the invariance group of the typical fiber of $E$. This secures the key idea that the typical fibers $\mathbb{C}^3\otimes\mathbb{C}^2\otimes \mathbb{C}$ play a role analogous to the one that $\RR^{4}$ plays for spacetime tensors, and the gauge groups, $SU(3)$ and $SU(2)\times U(1)$ are analogous to $O(3,1)$ (or $SO(3,1)$), viz. that they preserve the structure of the typical fibers.

Accordingly, in brief,  here is this paper\rq{}s  explanation for what I called the co-rotation of e.g. all the different quarks. The affine connection that defines their parallel transport is tied to the geometry of an `internal\rq{} space, a space which all these different fields share. This explanation is on a par with the reason given above for different kinds of tensors and spinors, e.g. $T_{\mu\nu}, X_{\mu}, A^{\mu}, \phi$, all being parallel-transported by the same affine (e.g. the Levi-Civita) connection: namely, that they are all tied to the tangent bundle  $TM$. Abstractly, we can represent the entire content of the standard model geometrically: the symmetry groups are merely the automorphisms of this geometric structure.  So, to repeat, here is this paper\rq{}s content, in slogan form: gauge transformations can be understood naturally as automorphisms of an internal geometric structure; and   the Ehresmann connection can be understood similarly to the Levi-Civita connection determining parallelism for tensor fields over spacetime.
 
 
 And here is how I will proceed: in Section \ref{sec:PFB_formalism} I will provide a unified view of principal fiber bundles, that will serve to explain parallel transport for both external, spacetime quantities, and for internal charges of the sort that appear in gauge theories. In Section \ref{sec:disa} I elaborate the two apparent disanalogies between general relativity and  gauge fields that I summarised above. In Sections \ref{sec:dispel_first} and \ref{sec:dispel_second}  I will dispel these disanalogies (respectively), along the lines just sketched: an enterprise which will involve showing how to the whole content of the SM consists of fields living on certain internal spaces.\footnote{ 
  In Appendix \ref{app:connection} I will provide more detail about the Ehresmann connection form}

 \section{Frame bundles and  parallel transport}\label{sec:PFB_formalism}\label{subsec:general_cons}

In Section \ref{subsec:PFB_ex}, I will provide a quick overview of fiber bundles in general, and principal fiber bundles  and their associated vector bundles in particular. The basic idea of a bundle is that it has internal spaces associates with each spacetime point---called fibers---and there is no canonical way to identify points in different fibers. In Section \ref{sec:parallel} we will see how an additional structure, the Ehresmann \emph{connection},  enables us to parallel transport elements of the fibers along  paths on spacetime. The application of this formalism to general relativity is briefly reviewed in Section \ref{sec:GR_gen}. 
 
\subsection{Fiber and frame bundles}\label{subsec:PFB_ex}


 The main idea of fiber bundles is that they are spaces that locally look like a product, i.e. they form a `bundle' of fibers over a base manifold (usually spacetime).
Let us denote fiber bundles by $E$; they are smooth manifolds that admit the action of a surjective projection $\pi:E\rightarrow M$ so that any point of $M$ has a neighborhood, $U\subset M$, such that $E$ is locally of the form $\pi^{-1}(U)\simeq U\times F$, where $F$, as above,  is some `fiber': a  space over each point of $M$ and in which the fields take their values, and  similarly for all subsets of $U$, which ensures that $\pi^{-1}(x)=F$. 
 But the isomorphism between $\pi^{-1}(U)$ and $U\times F$ is not unique,  which is why there is no canonical identification of elements of fibers over different points of spacetime. Each choice of isomorphism is called `a trivialization' of the bundle: it  is basically a coordinate system that makes the local product structure explicit. 
 
 So, to be explicit: $F$ is some space where   quantities in spacetime take their value, so it is usual to  denote a fiber bundle $E$ over $M$, with typical fiber $F$, by $(E,M, F)$. Fiber bundles admit \emph{sections}:
\begin{defi}[A section of a bundle] A field-configuration for  $(E, M, F)$ is called \textit{a section}, and it is a map $\kappa: M\rightarrow E$ such that $\pi\circ\kappa=\mathrm{Id}_M$.   We denote smooth sections like this by $\kappa\in C^\infty(E)$.
\end{defi} 
Sections replace the functions $\tilde\kappa:M\rightarrow F$, that we would employ if the fields  had a fixed, or  ``absolute''---i.e. spacetime independent---space of values.

There are many examples in which $F$ is a vector space. For instance,  a scalar field  takes values in $\RR$ or $\bb C$, whereas a more complicated field such as a  tangent  field or a spinor field, could take values in $\RR^4, \bb C^4$, etc. 

Alternatively, we could have bundles in which the typical fiber is isomorphic to a Lie group. These are known as \emph{principal fiber bundles.} 
A principal fibre bundle is, in short, just a manifold where some group $G$ acts, and whose equivalence classes under the group action correspond 1-1 to points of spacetime, $M$. We usually denote such a bundle by $(P, G, \pi, M)$. In detail:
\begin{defi}[a Principal fiber bundle]  is a smooth manifold $P$ that admits a smooth free action of a  (path-connected, semi-simple) Lie group, $G$: i.e.  there is a map $G\times P\rightarrow P$ with $(g,p)\mapsto g\cdot p$ for some left action $\cdot$ and such that for each $p\in{P}$, the isotropy group is the identity (i.e. $G_p:=\{g\in{G} ~|~ g\cdot p=p\}=\{e\}$).\end{defi}
  
Naturally, we construct a projection  $\pi:P\rightarrow{M}$ onto equivalence classes, given by  $p\sim{q}\Leftrightarrow{p=g\cdot{q}}$ for some $g\in{G}$. That is: the base space $M$ is the orbit space of $P$, $M=P/G$, with the quotient topology, i.e. it is characterized by an open and continuous $\pi:P\rightarrow M$. By definition, $G$ acts transitively on each fibre, i.e. on each orbit of the group.
Here, unlike in the general definition of a fiber bundle, we don't need to postulate the local product structure: $\pi^{-1}(U)\simeq U\times G$: it can be proven if the action of the group on $P$ is free. Nonetheless, an alternative construction of a principal fiber bundle postulates that it can be covered by such trivialisations into local products, with appropriate transition functions valued in $G$. In either case, it is clear that in the definition of a principal fiber bundle, the group $G$ is definitional: it is part of the fundamental structure of the bundle.\footnote{ It is somewhat confusing that a \textit{section of a vector bundle} is an entirely different object from the section of a principal bundle, which we will discuss below. So, for instance two different configurations of the electron field are two different sections of its vector bundle, and thus are not counted as `equivalent' in the way that two sections of a principal bundle are. And while a global section of $P$ exists iff the bundle is trivial, we can always find a global section of an associated bundle (cf. \cite[Theo. 5.7]{kobayashivol1}).\label{ftnt:section_triv} }

\paragraph{Example: the bundle of frames.}
 A  smooth tangent vector field is a section of the tangent bundle:  a smooth assignment of  elements of $TM$ over $M$.  In this case the familiar notation is  $X\in \mathfrak{X}(M)$ (instead of our more idiosyncratic $\kappa$), with $\pi:TM\rightarrow M$ mapping $X_x\in T_xM\rightarrow x\in M$. The tangent bundle   $TM$ \emph{locally}  has the form of a product space, $U\times F$, with $F\simeq \RR^4$.\footnote{Indeed, it is easy to see how the isomorphism is induced by taking the push-forward of a given chart $\phi: U\rightarrow \RR^4$.}  

 We can build a principal bundle as the set of all linear frames of $TM$, called `the frame bundle' (where `frame' means `basis of the tangent space $T_xM$'), written $L(TM)$.  The fibre over each point of the base space $M$ consists of all choices $\{\mathbf{e}_i(x)\}_{I=1, \cdots n}\in L(TM)$, of sets of spanning and linearly independent vectors (here the index $i$ enumerates the basis elements). So here $P=L(TM)$, with $p\in P$ a frame and  $\pi:P\rightarrow M$  the canonical projection onto the base point of the frame. Each element of $P=L(TM)$ is a linear isomorphism from $\RR^n$ into $TM$; a section of $P$ is a smooth assignement of a frame for $TM$ over each spacetime point.

\subsubsection{Associated bundle}\label{sec:associated}
Sticking to the example of the tangent bundle for now, in order to express a vector field as an element of an associated bundle, we take  a vector $X_x \in T_xM\simeq F=\RR^n$, where   according to a frame, $p=\{\mathbf{e}_i(x)\}\in L(TM)$, we write $X_x=a^i \mathbf{e}_i\in T_xM$ as the ordered quadruplet $(a^1, \cdots, a^n)\in \RR^n$.  We can rotate the frame  by a matrix $g^{ij}=\rho(g)$, where $\rho:G\rightarrow GL(\RR^n)$ is the matrix representative of the abstract group element, to obtain $\{g\mathbf{e}_i(x)\}\in L(TM)$. The components of $X_x$ will change accordingly, as $a^k\mapsto a^k g_{kl}^{-1}$. With the two transformations, we obtain the same vector: $a^k g_{kl}^{-1} g^{li}e_i=a^i e_i$. 

In the general case, if we write a doublet $(p,v)$ as, respectively, the frame and an element of the typical fiber, we want to identify $(gp, vg^{-1})$ (where we have simplified the notation for the action of the group to be just juxtaposition). So, given any vector bundle $(E,M, F)$, we have a principal $GL(n)$-bundle $P=L(E)$, and  we can express $E$ as an \emph{associated bundle}  $E\simeq L(E)\times_\rho F$ to $L(E)$, where:
\be L(E)\times_\rho F=L(E)\times F/\sim\quad \text{where}\quad (p,v)\sim  (gp, vg^{-1}),
\ee
and denote the equivalence classes with square brackets: $[p,v]\in   L(E)\times_\rho F$. 

 More generally, given any frame $p\in P=L(E)$,  there is a one-to-one map between its fibre  and the group $GL(F)$: we can use the group to go from any frame  to any other (at that same point), but there is no basis that canonically corresponds to $\mathbb{1}\in GL(F)$.  

 \subsubsection{Admissible bases and subgroups of $GL(n)$.}\label{sec:admissible}

 From what I have stated above, it might appear that we should only care about the general linear group, $GL(F)\simeq GL(n)$, where $n$ is the dimension of the typical fiber $F$. But for most uses, it is subgroups of the structure groups that are physically important.  For instance, for general relativity,  the structure group is $O(n)$ (or $SO(3,1)$)  acting on the orthonormal bases. So we in general do not consider the space of all frames, but only of some subset, called \emph{the admissible frames}.
 
How can we justify  the restriction of $\rho(G)$ to a subset of the most general group of transformations between frames? The restriction corresponds to the preservation of some added structure to $F$: the group reflects some intrinsic structure of the fibers. In other words, when $F$ is not just a bare vector space, but e.g. a normed vector space, we would like changes of basis to preserve this structure, e.g. the orthonormality of the basis vectors, and this restricts the bundle of linear frames to the appropriate sub-bundle of \emph{admissible frames}; and the transformations between admissible frames belongs to a reduced structure group.

We give an example of this in Section \ref{sec:GR_gen}, where, requiring $F$ to be endowed with an inner product, reduces the structure group to $G=O(n)$; similarly, $SO(n)$ adds an orientation to $F$; and $G=U(n)$ corresponds to a complex vector space structure and a Hermitean inner product; and $G=SU(n)$ adds an orientation. The moral is that the added structure on $F$ induces an added structure on  the associated vector bundle if and only if the  transformation group $G\subset GL(n)$ preserves that added structure. The symmetry groups are groups of automorphisms of a geometric structure at each point of spacetime. 

In sum,  starting with a given vector bundle $E$, the frame bundle  $L(E)$ (formed by the bases of $E_x$ for each $x\in M$) is a principal bundle $P'$ with structure group  $GL(F)$. And since $P'\simeq L(E)=L(P\times_\rho F)$, we can see any other $P$ whose $G$ has a faithful representation on $F$  as a sub-bundle of $L(P\times_\rho F)$ corresponding  to a subset of frames of $L(P\times_\rho F)$ related by $\rho(G)$ (the admissible frames). The main advantage of seeing a principle bundle $P$ as a bundle of frames of a vector bundle $E$ is that \emph{{the gauge group is no longer postulated as fundamental; it acquires meaning as the invariance group of the typical fiber of $E$}}. 



Here $P$  does not   represent states of matter: that role is reserved for sections of different vector bundles $E$; nor does $P$ represent points or vectors in spacetime.  Instead, it represents bases for the spaces in which particle fields take their value. And gauge transformations acquire a rather mundane explanation: they correspond to spacetime-dependent changes of bases of the inner vector spaces of the vector bundles.\footnote{These are  \emph{passive} gauge tranformations. And we can interpret \emph{active} gauge transformations as active rotations of these spaces (in which the bases stay the same but the components of the vectors rotate). This is just the difference between an active vertical automorphism of the bundle and its corresponding gauge transformation between sections.\label{ftnt:passive_active} } Similarly, the principal connection---the notion of horizontality in the PFB with structure group $G$ to be discussed in Section \ref{sec:parallel} below---induces a notion of parallel transport in the  associated vector bundle. 
\citet[p. 2404]{Weatherall2016_YMGR} writes: 
 \begin{quote}
 We are thus led
to a picture on which we represent matter by sections of certain vector bundles (with additional structure), and the principal bundles of Yang–Mills theory represent various
possible bases for those vector bundles. 
[...]
 these considerations lead to a deflationary attitude towards
notions related to ``gauge'': a choice of gauge is just a choice of frame field relative
to which some geometrically invariant objects [...] may be represented, analogously to how geometrical objects may be
represented in local coordinates.\end{quote}


\subsection{Parallel transport.}\label{sec:parallel}

 Here is how to visualise parallel transport in this principal fiber bundle picture. Directions transversal to the fiber will relate frames over neighbouring points of $M$; they will tell us which basis over $x+\delta x$ correspond to a chosen basis over $x\in M$. Imagining the manifold $M$ to lie horizontally on the page, we think of the fibers as vertical, and, on $P$, we  dub as \textit{horizontal} a preferred set of directions transversal to the fibers, that we take as a preferential link between the frames on neighbouring fibers. The horizontal space at a point is isomorphic to the tangent space of the base manifold under that point: $H_p\simeq T_{\pi(p)}M$. So, the vertical spaces---the fibers---are part of the basic structure of the principal bundle, but a preferred choice of a transversal distribution---called horizontal---is not (it is usually a dynamical part of field theories).

So parallel transport is described by a distribution of tangent subspaces transversal to the fibers of a principal bundle: $P\ni p\mapsto H_p\subset T_pP$. 
We can define this distribution by the use of a \emph{connection-form}, $\omega$, which, in the case of frames, tells us how a frame gets transported to a neighboring spacetime point (for completeness, I provide a more detailed mathematical exposition of the connection form in Appendix \ref{app:connection}).

 For $(E, M, F)$,  we describe $L(E)$ using its frames   $\mathcal{B}=\{{\mathbf e}_i\}_{i=1}^k$ and their algebraic duals  $\mathcal{B}^*=\{{\mathbf e}^i\}_{i=1}^k$.  Since a linear transformations of $F$ is an element of $F^*\otimes F$, we write the connection as:
\be\label{eq:omega_tetrad}\omega=\omega_i^j\otimes {\mathbf e}^i\otimes {\mathbf e}_j\ee
 The matrix of one-forms on $P$, $\omega_i^j$ is the connection-form of Definition \ref{def:Ehres}, defined in terms of a representation $\rho$ of $\mathfrak{g}$ on $F$.\footnote{I am sacrificing rigour for fluency here, and will do so for the remainder of this Section. For the full account, consult e.g. \cite{kobayashivol1}.}

To calculate the parallel transport of sections of $(E, M, F)$,  this definition is employed by choosing a particular trivialisation $\sigma$, or section of $P$---a local, smooth assignment of frames to spacetime points---and pulling back  the connection-form to this section. In terms of this section, the connection form is expressed as a matrix of one-forms on the space of vectors of $M$.

 Thus, on a section $\sigma$ of the frame bundle, parallel transport of any associated bundle is specified by how the frame rotates along any given direction, i.e. by the covariant external derivative of the frame:
\be \D{{\mathbf e}_j}=\omega^i_{\phantom{i}j}\otimes {\mathbf e}_i,\ee 
where now,with a slight abuse of notation, $\omega^i_{\phantom{i}j}$ is  matrix of one forms on $U\subset M$, the domain of the section of $P$ (see e.g. \cite[Ch. 5]{kobayashivol1}). Under a change of local section, or trivialisation, $\sigma\rightarrow \sigma\rq{}$, described by a function  $h:U\rightarrow G$, $\omega$ changes in the familiar way: 
\be\label{eq:gauge_transf}\omega\rq{}=h\omega h^{-1}+h^{-1}\d h. 
\ee

For some associated vector field, i.e. a section of $(E, M, F)$, we  write $s=s^i{\mathbf e}_i$, according to the chosen Section of $P$,  and define the external covariant derivative (or gradient) of $s$, as: 
\be\label{eq:cov_dev} \D {s}=\d s^j\otimes{{\mathbf e}_j}+s^i\omega^i_{\phantom{i}j}\otimes{e}_j.\ee

\subsection{General relativity and frame bundles}\label{sec:GR}\label{sec:GR_gen}

Here I will present the basics of the tetrad formalism for general relativity, which is very similar to that of Yang-Mills theory; both are easily understood in terms of the frame bundles.

The Palatini formalism for general relativity no longer works directly with an affine connection  derived from the metric, but rather with the aforementioned $SO(3,1)$ Ehresmann connection associated with  the frame bundle $L(TM)$. Briefly, the formalism is defined as follows. 

Given $L(TM)$,  I will first endow $M$ with a Lorentzian metric, $\langle\cdot, \cdot\rangle$ and the fibers $F\simeq \RR^4$ with a Minkowski inner product, which I will call $(\cdot,\cdot)$. Now we can induce a subbundle $P\subset L(TM)$  by requiring that $p\in P$ iff given $u, v\in \RR^{4}$ and:
\be\label{eq:ortho_abs} (u,v)=\langle pu, pv\rangle.\ee
So a basis or frame is admissible iff it is orthonormal with respect to $\langle\cdot, \cdot\rangle$.  This construction yields a principal fiber bundle with structure group $G=O(3.1)$; similarly, $SO(3,1)$ adds an orientation to $F$, and so on.\footnote{Here I have given a consistency condition, but one could have equally well defined a metric from the class of admissible frames and the inner product of the fiber, or an inner product on the fibers by the metric and class of admissible frames. For instance, take again $p\in P$ as a linear isomorphism from $\RR^{4}$ to $T_{\pi(p)}M$: each $p$ defines an orthonormal frame. Define an inner product on $TM$ , for $X,Y\in T_xM$ as \be (p^{-1}X, p^{-1}Y)=\langle X, Y\rangle.\ee Invariance of $(\cdot,\cdot)$ by $O(1,3)$ implies the inner product is independent of which basis $p\in \pi^{-1}(x)$ we take.  
In more detail and generality, define $P\times_\rho F$ as the equivalence class for the doublet $(p, v)\in P\times F$  with  $(p, v)\sim (g\cdot p, \rho(g^{-1})v)$. Suppose  that $F$ is a Riemannian vector space, with metric $\langle\cdot,\cdot\rangle$. We can  induce a metric in $P_F=P\times_GF$  defining, for any  $p$ and $v,v'\in F$:
$\langle[p,v],[p,v']\rangle:=\langle v,v'\rangle$.
To be well-defined, we must have:
$$\langle[p,v],[p,v']\rangle=\langle[g\cdot p,\rho(g^{-1})v],[g\cdot p,\rho(g^{-1})v]\rangle=\langle \rho(g^{-1})v,\rho(g^{-1})v'\rangle:$$
which is true only if the action of the group on $F$ is orthogonal with respect to the metric.}

In a coordinate chart for $M$, i.e. a local diffeomorphism $\phi:M\supset U\rightarrow \RR^4$ with coordinates $x^\mu$,  we write ${\mathbf e}_i=e_i^\mu\frac{\partial}{\partial x^\mu}$, with Latin indices reserved for the elements of the frames and Greek indices for the coordinate basis. The reason that I call the typical fibers of vector bundles associated to the bundle of frames $L(TM)$ `external\rq{} is that the tetrad ties, or `solders\rq{} the typical fibers to the tangent space, as indicated by the mixed indices that it carries. The curvature $\Omega$ (cf. equation \eqref{eq:Omega}) and the torsion $\Theta$ of the (dual) tetrad are called \emph{Cartan\rq{}s} structural equations: 
\be\label{eq:torsion}\Theta:=\d {\mathbf e}^i+\omega^i_{\phantom{i}j}\wedge {\mathbf e}^i;\quad \Omega^i_j:=\d \omega^i_{\phantom{i}j} +\omega^i_{\phantom{i}k}\wedge\omega^k_{\phantom{i}j}
\ee

Calling the coordinate expression of the metric $\langle \cdot, \cdot\rangle=g_{\mu\nu} \d x^\mu\d x^\nu$, we have: 
\be\label{eq:metric_viel}g_{\mu\nu} e_i^\mu e_j^\nu=\eta_{ij},\quad \text{and}\quad  g^{\mu\nu}=\eta^{ij}e_i^\mu e_j^\nu
\ee
where $\eta_{ij}$ is the  canonical Minkowski inner product on $\RR^4$, the diagonal metric $(-1,1,1,1)$, with which one raises and lowers the frame indices. Equation \eqref{eq:metric_viel} subsumes the compatibility relation between fiber inner product and manifold metric, \eqref{eq:ortho_abs}. 

We get two restrictions by applying the covariant derivative to the equations \eqref{eq:metric_viel}: the first applies to all connections that preserve the fiber structure, and it defines the action of the external covariant derivative $\D$ of \eqref{eq:cov_dev} on the one-forms $\mathbf{e}^i$ which is torsion-free ($\Theta=0$ in \eqref{eq:torsion}), 
 and the second applies to those connections that are compatible with the spacetime metric, ensuring the matrix $\omega^i_{\phantom{i}j}$ is anti-symmetric (so that it represents the Lie algebra $\mathfrak{so}(3,1)$); jointly these conditions give the Levi-Civita connection.  The extension of $\D$ from \eqref{eq:cov_dev} to arbitray vector-valued forms is then determined  by anti-linearity and its action on the tetrad basis and dual basis (see e.g. \cite[Ch. 3.4b and Appendix B]{Wald_book} and \cite[Ch. 14.3-14.6]{MTW}). 

The Riemann curvature can be expressed simply as the matrix-valued two-form $\Omega$ in \eqref{eq:Omega}: 
\be \Omega^i_{\phantom{i}j}=R^i_{\phantom{i}jk\ell} {\mathbf e}^k \wedge {\mathbf e}^\ell.\ee
In this form,  the Bianchi identity is a consequence of the nilpotency of the external derivative, and can be written simply (omitting the matrix indices) as: $\D\Omega=0$.

It is possible to show that the Einstein equations and the metric compatibility emerge from a variation in both the tetrad and the connection form of  the very simple Palatini action functional (see e.g. \citep{Burton1998} and references therein):
\be \int_M \d^4 x |e| e^\mu_i e^\nu_j \Omega_{\mu\nu}^{\phantom{\mu\nu}ij}
\ee
where $\d^4 x |e|$ is the volume element ${\mathbf e}^0\wedge {\mathbf e}^1\wedge {\mathbf e}^2 \wedge {\mathbf e}^3$. 

\section{Disanalogies?}\label{sec:disa}

Focussing on our understanding of the role of frames and parallel transport, I will now elaborate the two possible disanalogies  between the general relativistic and the gauge theoretic cases, that hinder explanation in the gauge case. The first disanalogy will be described in detail in Section \ref{sec:first_disa}: it is about the necessity of the use of frames in order to explicitly describe parallel transport,  of either tensor bundles or vector bundles associated to the particles of the SM.  The second disanalogy will be described in Section \ref{sec:second_disa}: it is about how parallel transport of elements of the tangent bundle automatically induce, or geometrically force, a unique notion of parallel transport on other tensor bundles, so that they all march in step; whereas there is apparently no similar geometric foundation for the co-rotation of elements of different vector bundles in the standard model, even those that fall under the same interaction group.

We can extract these two disanalogies from \cite[p. 2404]{Weatherall2016_YMGR}\rq{}s description (in the opposite order than I have): 
\begin{quote}
in general relativity, all of the bundles associated to the frame bundle ---that is, the tangent bundle, the cotangent bundle, and bundles of tensors acting on them--- can be defined directly in terms of tensor products of the tangent bundle and cotangent bundle, which in turn can be defined directly in terms of tangent vectors and linear functionals. This method of constructing the bundles  allows one to take a covariant derivative operator on the tangent bundle and [second disanalogy:] \emph{immediately induce a corresponding derivative operator on all of the other vector bundles one cares about}, [first disanalogy:] \emph{without ever mentioning that they all correspond to a single principal connection on the frame bundle}. [my italics]
\end{quote}


\subsection{The first disanalogy: are frames unnecessary in the case of spacetime?}\label{sec:first_disa}
 Using frame bundles over $TM$, the tetrad formalism (as described in Section \ref{sec:GR_gen}) gives a concise and self-contained description  of spacetime geometry and the Einstein equations, which does not take the Levi-Civita affine connection $\nabla$ as the fundamental object defining parallel transport.  Instead, in this description Einstein’s equations govern the curvature of the principal connection on a principal bundle over spacetime, and the covariant derivative operators  on  associated bundles are induced by that connection.  And although the formalism introduces an  `internal\rq{} $SO(3,1)$ symmetry to the theory which might seem superfluous, in a more rigorous mathematical sense no extra structure has been added.\footnote{For any manifold $M$ gives rise, in a canonical way, to its frame bundle $L(TM)$, which, having fixed an isomorphism between $L(TM)\times_\rho \RR^4$ and $TM$, tie $\omega$ to $\nabla$, the covariant derivative acting on all the tensor bundles over $M$.  The sense in which there is no extra structure is that the group of automorphisms of the frame bundle  that preserve the isomorphism between $L(TM)\times_\rho \RR^4$ and $TM$
is canonically isomorphic to $Diff(M)$, the diffeomorphism group of $M$ (cf. \cite[p. 2399]{Weatherall2016_YMGR}). }  Lastly, beyond mathematical simplicity and beauty, the formalism is extremely powerful. For instance, according to the authoritative \citet[p. 351]{MTW}, \lq\lq{}Ordinarily, equation (14.18) [our \eqref{eq:Omega}, which describes curvature in terms of the tetrads and the spin connection] surpasses in efficiency every other known method for calculating the curvature\rq\rq{}; in his derivation of the Schwarschild metric, \citet[Ch. 6.1]{Wald_book} uses the tetrad approach; etc.

Notwithstanding  structural equivalence, mathematical simplicity, and utility of the frame description, there is a perfectly valid alternative to describing parallel transport of tensors, via the Levi-Civita connection $\nabla$, that does not mention frames or $SO(3,1)$. Again,  \citet[p. 2399]{Weatherall2016_YMGR} describes   the general relativistic case as follows: 
\begin{quote}
It is this object [$\nabla$] that determines the trajectories of \textit{massive bodies and the dynamics of (tangent
valued) matter fields}. Moreover, this derivative operator (and its associated curvature)
may be fully and invariantly characterized without ever mentioning frames, gauges,
connection coefficients, or anything of the sort. The frame bundle merely provides
an alternative---and for many purposes, less attractive---way of encoding information
about this derivative operator.
\end{quote}
So the idea of the disanalogy is that the principal bundle and Ehresmann connection are \emph{completely}  unnecessary in order to explain the co-rotation of sections of different tensor bundles, and they are necessary---but merely `auxiliary\rq{}---in order to explain the co-rotation of sections of different associated vector bundles in the gauge case.

But my italics in this quote point to a shortcoming of the description of parallel transport of sections of $TM$ via $\nabla$, relative to the description via frames, that I will exploit in Section \ref{sec:spinors}. Namely, in modern physics matter fields are \emph{spinorial}. Spinors are more directly related to the Lorentz group $SO(3,1)$, and are usually described in terms of frames.  So spinors provide a very tight analogy to the gauge case: it is convenient, but not strictly necessary, to express their properties in terms of bundles of frames.

\subsection{The second disanalogy: the universality of parallel transport}\label{sec:second_disa}

  Put the first disanalogy---about the necessity of using frames---aside, as it will be dealt with separately. The second disanalogy is about a unified explanation for parallel transport of different vector bundles marching-in-step, or co-rotating.

Again, recall that in the case of spacetime,  
  because tensor bundles are built up from  $TM$, its dual---$T^*M$, which consists of linear functionals on $TM$---and their tensor products, it is immediate that parallel transport for all  tensor bundles would `march in step'.  Any structure that determines parallel transport on $TM$ will determine it also on arbitrary tensor bundles formed from $TM$, e.g. $TM\otimes \cdots \otimes TM\otimes T^*M\otimes \cdots \otimes T^*M$. 
Thus, in general relativity, once we have interpreted the  connection in terms of a basis of $TM$, we are obliged to use the same connection for other tensor bundles.\footnote{This kind of construction gives a geometric underpinning to the classification of tensor bundles as \emph{natural bundles}. For our purposes,  natural bundles are those bundles over $M$ such that any diffeomorphism of $M$ has a unique lift to sections of those bundles. See e.g. \cite[Ch. 4]{Kolar_book} and \cite[Ch. 4]{Fatibene2003} (and see \cite[Ch. 5]{Kolar_book} for a proof that any natural bundle can be canonically associated to a (possibly higher-order, jet) bundle of frames). Similarly, for \emph{gauge-natural bundles}, the principal bundle $P$ and its automorphisms play the analogous roles of $M$ and its diffeomorphisms: now the automorphisms of $P$ admit a unique lift to sections of gauge natural bundles.\label{ftnt:natural}} 

In the gauge theoretic case, we can still think of the same principal connection as encoding the parallel transport of the different matter fields, but it is the matter fields that are physically significant. Here is \cite[p. 2404]{Weatherall2016_YMGR} again: 
\begin{quote}
from this point of view the standard terminology is misleadingly backward: it is the vector
bundles that matter, and for some purposes, one might also introduce the associated
principal bundles. The real physical significance should be attached to (sections of)
the  [...] vector bundles and the covariant derivative operators acting on them.
[...]
But this does not quite mean one can forget about the principal bundles altogether 
[...] the principal bundles in Yang–Mills theory coordinate frames across different vector bundles. And this coordination of frames allows one to make precise the senses in which (1) vectors in different vector spaces might be constant by the same standard of constancy and (2) different vector bundles might have the same (dynamical) curvature. 
\end{quote}

But why should the frames be coordinated? 

To recall:  in the gauge theoretic case, the fundamental forces are associated to Lie groups, and each field that interacts via such  a force lives in a vector bundle that admits an action of the corresponding group. 
But each charged matter field inhabits its own vector bundle, which means that  the connection acts on many different vector spaces. Thus  it would seem that we  cannot interpret the gauge connection in the same way as the Levi-Civita connection: as encoding parallel transport for a fundamental type of internal (external, for the Levi-Civita) geometry, and then have that automatically induce a notion of parallel transport for all of the vector (resp. tensor) bundles.   In the gauge theoretic case, we could conceivably use different principal Ehresmann connections to describe the parallel transport of different particle types. What seems possible in the gauge case is a geometric impossibility in the case of different tensor fields.\footnote{In footnote \ref{ftnt:natural}, I mentioned the concepts of natural and gauge-natural bundles. The disanalogy pointed to here persists also in that terminology: with the exception of  force-carrying bosons, the particle fields  of the SM don\rq{}t get their `naturalness\rq{} in the same way as tensorial fields: for we do not construct them from some underlying geometric structure, but assume that they are sections of vector bundles associated to $P$.  }

The extant literature provides little explanation for the foreclosure of such a possibility in gauge theory.  Weatherall goes furthest, and still, besides the clarification in the quote above, he offers only the metaphor:  \lq\lq{}principal bundles are auxiliary [...] in  the sense in which a coach is auxiliary to the players on the field." 
 The way to dispel this putative disanalogy to the gravitational case will be  to construe all the vector bundles of the SM as being built from the same underpinning internal geometry, in analogy to how tensor bundles are built from $TM$. The gauge group emerges simply as the invariance group of this geometry.  I will pursue this argument in Section \ref{sec:problems}. 

By dispeling these two disanalogies, I will show that gauge transformations can be understood naturally as automorphisms of an internal geometric structure; and   the Ehresmann connection can be understood similarly to an affine connection  over spacetime.

\section{How to dispel the first disanalogy}\label{sec:dispel_first}

Now I will dispel the first disanalogy. I have two lines of arguments, to be elaborated in Sections \ref{sec:spinors} and \ref{sec:parallelism}, respectively.

In Section \ref{sec:spinors} I will give a very brief and very narrow introduction to spinors, one of the fundamental entities in the standard model, which has, remarkably, received little attention in the literature on the foundations of spacetime theories. Spinors are intimately tied to $SO(3,1)$ and with the bundle of frames, and yet have their parallel transport determined by $\nabla$, the Levi-Civita connection. They are, in an important sense, tightly analogous to sections of vector bundles associated to a  principal bundle with gauge group $G$. 

 In Section \ref{sec:parallelism}, I argue that, though parallel transport as described in a principal fiber bundle may employ frames, it can be abstractly characterised in a frame-independent manner; so here too, the parallel transport of external and the internal spaces is tightly analogous.  

\subsection{Spinors}\label{sec:spinors}
 
  There are two important reasons for this Section\rq{}s exposition. The first, to be discussed in Section \ref{sec:intro_spin}, is that it helps to dispel the first disanalogy, described in Section \ref{sec:first_disa}. That is because  spinors are intimately tied to the Lorentz group $SO(3,1)$ (through its double cover). Thus we have good reason to think of parallel transport in curved spacetimes as described, in this more general sense, by a principal fiber bundle with structure group $SO(3,1)$. This description admits of changes of frames as a kind of gauge transformation, and so, \emph{contra \cite{Weatherall2016_YMGR},} is in  a tighter analogy to the descriptions of the principal bundle, parallel transport, etc, in the case of gauge fields. 
 
 The second reason, to be discussed in Section \ref{sec:geom_spin}, falls slightly outside of this Section\rq{}s main theme, of dispelling the first analogy. Namely, it is that spinors illustrate the possibly elaborate constructions of vector bundles from the geometry of the fundamental typical fiber---in this case, an \emph{oriented} tangent space, endowed with an inner product.  So they illustrate, in the case of the `external\rq{} spaces, my claims about gauge: that different particle fields co-rotate under gauge transformations and parallel transport \emph{because} they share, or are constructed from, the same underpinning `internal\rq{} space. Even if this construction might seem at time  circuitous.\footnote{In fairness, such construction is much more circuitous in the case of spinors than in the case of internal geometries that appear in the SM. } In other words,  the significance of $SO(3,1)$ for the parallel transport of spinors stems from geometrical aspects of $TM$; although frame bundles are useful in characterising spinors, they do not carry fundamental physical significance.

\subsubsection{A brief introduction to spinors}\label{sec:intro_spin}

This Section\rq{}s introduction follows \cite[Ch. 13]{Wald_book}.

 In brief: mathematically, a spinor at $x\in M$ is an ordered pair of complex
numbers associated with an orthonormal basis of the tangent space  which transforms
in a specified way under a continuous change of basis. This definition suffices for the purposes of this paper. But of course, I will elaborate. 

In introductory studies of quantum mechanics at the undergraduate level, spinors are typically encountered within the framework of non-relativistic quantum mechanics, particularly in the treatment of spin angular momentum. This initial encounter often leads students to associate spinors exclusively with spin, a connection reinforced by the term `spinor' itself. However, let us refrain from any relation to quantum mechanics. Instead, consider the term `spinor' as a broader concept akin to `vector' or `tensor'. The basic features of a spinor is that it is related to the null cone, and to peculiar behavior under $2\pi$ rotations.  Thus, the numerical values of a spinor in a given orthonormal
basis cannot be directly physically measurable since it has two possible distinct
values in that basis. Clearly, this contrasts sharply with the analogous transformation
laws for ordinary tensors.  However, real bilinear products of spinors and complex conjugate
spinors may be identified with ordinary vectors and thus have a direct physical
interpretation. Indeed every null vector can be expressed as the tensor product of a spinor and its complex conjugate. In this sense, a spinor may be viewed as a \lq\lq{}square
root\rq\rq{} of a null vector. I\rq{}ll elaborate the geometrical meaning of spinors in Section \ref{sec:geom_spin}.\footnote{One should heed renowned mathematician Michael Atiyah\rq{}s warning: \lq\lq{}No one fully understands spinors. Their algebra is formally understood but their general significance is mysterious. In some sense they describe the "square root" of geometry and, just as understanding the square root of -1 took centuries, the same might be true of spinors.\lq\lq{} (as quoted in \cite{Farmelo2010}).\label{ftnt:Atyiah}}

The basic idea for constructing spinor fields in a curved spacetime $\langle M, g_{ab}\rangle$ is to start with $(P_o, SO(3,1), \pi_o, M)$, where $P_o\subset L(TM)$ is the principal fiber bundle of oriented, time oriented orthonormal frames, as described in Section \ref{subsec:PFB_ex}, so that each fiber is isomorphic to the proper Lorentz group. Then one \lq\lq{}unwraps\rq\rq{} each fibre to produce $(P, \text{Spin}(3,1), \pi, M)$, the spin bundle over $M$. In other words, the spin bundle replaces $SO(3,1)$ by its double cover, $\text{Spin}(3,1)$  as its structure group (which has the same Lie algebra, $\mathfrak{so}(3,1)$).\footnote{There are major topological obstructions to having a spin bundle over a general manifold $M$. A necessary and sufficient condition for $M$ to admit a spin structure is for its second Stiefel-Whitney class to vanish: for open, simply connected manifolds, this means that they are parallelizable (the tangent bundle is trivial).  However, once one considers internal gauge groups with the appropriate $\bb Z_2$ action, one can form spin bundles over more general manifolds (though one may lose uniqueness of these structures): see \cite{Isham_spin}.} 

The spinor bundle then is constructed as a fiber bundle associated to this principal bundle with fiber $F = \bb C^2$, with  the natural action of $\text{Spin}(3,1)\simeq SL(2, \bb C)$ on $\bb C^2$. A spinor at $x$ is an ordered pair of complex numbers which transforms by the natural representation of $SL(2, \bb C)$ under a change of orthonormal frame. 
In this kind of description,   spinor space is an internal space, but its changes of frame are driven by those of the spacetime manifold. In the next Section we will see how spinors are in fact part of the `external\rq{} space.
  
In this description, the characterisation of parallel transport is straightforward. Let $\omega$ be the Ehresmann connection given in \eqref{eq:omega_tetrad}, for a given section of the bundle $P_o$, and let $\phi$ be the projection from $P$ to $P_o$, induced by the projection from $\text{Spin}(3,1)$ to $SO(3,1)$.  
\[
\begin{tikzcd}
 P\arrow[dr,"\pi" ]\arrow[rr,"\phi"]&&  P_o\arrow[dl,"\pi_o"]\\
&M
\end{tikzcd}
\]
Then we can pull back the connection on $P_o$ to $P$ via $\phi$, and this defines parallel transport of spinors from parallel transport according to an $SO(3,1)$ connection on the bundle of frames. 

But in order to complete my argument, I would like to show that the spinor\rq{}s parallel transport marches in step with the parallel transport of other spacetime quantities---such as tensors---\emph{because}, in some sense, they both exist as spacetime geometrical quantities: we will do this in the next Section.\footnote{
I have tried to leave out of this brief exposition the many intricacies related to Clifford algebras, complexifications, etc. But if one thinks of complexification as a natural operation in geometry, using an orthonormal basis, i.e. a section of the $SO(3,1)$ frame bundle over $TM$,  $\mathcal{B}=\{{\mathbf e}_i\}_{i=1}^k$, we could define a pair of Weyl spinors as:
 \be\label{eq:Weyl} {\text{\bf w}}_j = \frac{1}{\sqrt{2}} \left({\mathbf e}_{2j} + i{\mathbf e}_{2j+1}\right); \quad {\text{\bf w}}_j^* = \frac{1}{\sqrt{2}} \left({\mathbf e}_{2j} - i{\mathbf e}_{2j+1}\right),\ee
 which, when properly examined in light of the Clifford algebra,  are naturally anti-commuting: ${\text{\bf w}}_j{\text{\bf w}}_m = -{\text{\bf w}}_m{\text{\bf w}}_j$; this is the property that realises the Pauli exclusion principle.
In relation to this interpretation and to Atiyah\rq{}s quote in footnote \ref{ftnt:Atyiah}, it is appropriate  to point out that, later in life (in a 2013 lecture at IHES), Atyiah said: \lq\lq{}spinors are the square root of geometry because complex geometry is the square root of real geometry\rq\rq{}. \label{ftnt:Weyl} } 
 
 Summing up: in order to describe spin bundles and their sections, which are tightly related to geometric structures that are preserved by $SO(3,1)$, we employed the bundle of orthonormal frames of $TM$. And as to parallel transport for spinors:  it is uniquely determined from the parallel transport of tensor fields, obtained from the Levi-Civita connection $\nabla$, or from the Ehresmann connection $\omega$ for the bundle of $SO(3,1)$-admissible frames (as discussed in Section \ref{sec:admissible}). From the perspective of tensor fields, a spinor will be said to be parallelly transported along a curve iff its `null flag\rq{} (cf. next Section)  is parallelly transported and its sign changes continuously. 
 
  In the next Section I will provide a brief overview of the standard geometrical interpretation of spinors, as \emph{null flags}, without the use of complexification or frame bundles.

\subsubsection{Geometric interpretation of classical spinors}\label{sec:geom_spin}

We saw in Section \ref{sec:second_disa} that the parallel transport for tensor bundles marches in step because they are constructed from the same fundamental geometry of $TM$, or still, from the same geometric fiber, $T_xM\simeq \RR^4$. But remember that we have endowed $TM$ with more than just affine and differentiable structure: it has a Lorentzian inner product, and possibly, even an orientation. We could use this structure to build more complicated vector bundles, and  parallel transport for $TM$ will then geometrically determine  parallel transport of sections of those bundles: the spin bundle is just such an example. I will argue that the same is true for gauge theories; there too,  there are many types of vector bundles that can be constructed from the same underpinning geometric structure. Since it is important for the main thesis of this paper that  parallel transport of different fields marches in step because they share a certain geometric space, I will now provide a geometric interpretation for spinors that does not rely on the frame bundle (as introduced in  \citep{penrose1968battelle}, but see e.g. \cite[Ch. 1]{Penrose1984} or \cite{steane2013} for a gentle introduction).

A (rank 1) Weyl spinor can be geometrically interpreted  as a null vector endowed with additional characteristics: a 'flag' identifying a plane in space containing the vector---representing overall `phase\rq{} by its orientation and `scale\rq{} by its size---and an overall sign, representing left or right-handedness. A null vector, $u$, defining the direction of the flag-pole, can be thought of as a point in a two-sphere---the light-sphere at time $t=T$ emanating from an event at $t=0$. 
Then the flag can be thought of as a vector tangent to this sphere and originating on $u$.   When such a spinor describes a particle's property, it lies on the future light-cone of the events where the particle is defined. But the past light-cone could also be associated to the same $u$. Indeed, the double-cover of the spin group can be thought of as a certain redundancy in terms of the above definition, for given the event at $t=0$, there is an identical light-sphere at $t=-T$. This ambiguity is associated to the chirality of the spinor and to its representation of matter or anti-matter:  e.g. `left-handed' means that the spin is in the opposite direction of the momentum; that could be interpreted as seeing the rotation of the spin from behind. (\emph{Mutatis mutandis} for the right-handed spinor). Thus, in standard presentations, matter is left-handed while antimatter is right-handed (massless spinors are all left-handed). A Dirac spinor has double the components of a Weyl spinor, and thus includes in principle both a right-handed and a left-handed part. 

As I mentioned in Section \ref{sec:intro_spin}, the unique characteristic of a spinor is its behavior under rotation: similar to a vector, the direction of a spinor changes under rotation, with the flag rotating alongside as if rigidly attached to a `flag pole'. Notably, while a rotation about the axis defined by the flagpole leaves a vector pointing in the same direction unaffected, it impacts the spinor due to its effect on the flag: under rotations of $2\pi$ the flag will be in the opposite direction from which it started, coming to its original position only under rotations of $4\pi$. There are in the literature several attempts at interpreting this feature geometrically (e.g. `Dirac\rq{}s strings\rq{}), but that won\rq{}t be necessary here. 

\begin{figure}[h!]
\center
\includegraphics[width=0.3\textwidth]{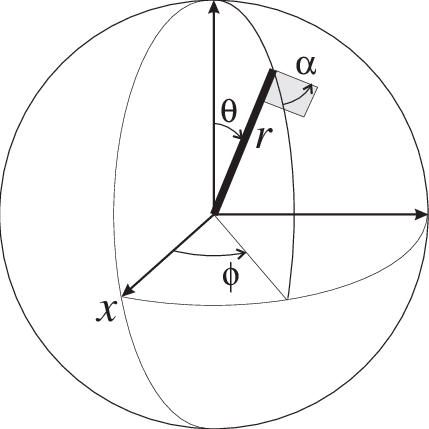}
\caption{ A visualisation of the flagpole, the flag, and the parameters determining a spinor of rank 1, taken, with permission, from \citep{steane2013}. }\label{fig:spinor}
\end{figure}

Thus, in order to define a spinor state, one needs to provide four real parameters along with a sign, denoted as \( r, \theta, \phi, \alpha \).  See figure \ref{fig:spinor}. 
 During rotation, the spinor's magnitude remains constant while the angles \( \theta, \phi, \alpha \) change. In the flag analogy, both the flagpole and flag move together as a single unit, allowing us to determine how \( \alpha \) changes with \( \theta \) and \( \phi \). To express the equations governing rotation effects conveniently, we combine the four parameters into two complex numbers (as obtained through a stereographic projection from the spheres to the complex planes). This is the $\bb C^2$ that describes the typical fiber of the Weyl spinor (see also equation \eqref{eq:Weyl} and footnote \ref{ftnt:Weyl}). A Dirac spinor would correspond to an element of $\bb C^4$, which can include both the left and right-handed Weyl spinors. 

And, as I said in the previous Section, a  spinor will be said to be parallelly transported along a curve if its null flag  is parallelly transported and its sign doesn\rq{}t change discontinuously.\footnote{Were we to drop the no-torsion condition, that ensures the inner product of the orthonormal frames is covariantly constant (as discussed in Section \ref{sec:GR_gen}), the parallel transport of a null flag (through its component spinors) would not retain the form of a null flag. } This suffices for my geometric explanation of parallel transport of spinors marching in step with tensors: they share the same geometry of spacetime.

This concludes the spinors\rq{} role in my overall argument about gauge theory. They  are very similar to gauge fields in that, in order to more economically describe them, one often resorts to frames and to the structure group. Nonetheless, as with gauge fields, the spinor  is geometrical: it \emph{can be} described independently of the structure group, just not---in the case of spinors---in a completely straightforward way.

\subsection{Parallelism without frames}\label{sec:parallelism}

Strictly speaking, this Section does not contradict the letter of any argument of \cite{Weatherall2016_YMGR}, who admits that a description of the Levi-Civita connection as an affine covariant derivative $\nabla$ is equivalent to that as an Ehresmann  $SO(3,1)$-connection $\omega$ in the bundle of frames. Nonetheless, it is useful to clarify in which way frames are necessary or unnecessary, and how the frame description of the Ehresmann connection corresponds to a frame-independent notion of parallelism for the typical fibers. This is important if we want to describe the gauge groups as secondary, as being the group of automorphisms of some internal geometric structure, which is what would come first.

Recall the putative disanalogy:  Lie groups seem to appear \emph{explicitly} in the principal fiber bundles encoding the parallel transport of particles fields, whereas these groups need not be invoked for the parallel transport of tensor fields in spacetime. But as I argued in Section \ref{sec:admissible},  Lie groups of a principal bundle seen as a bundle of frames of a vector bundle reflect the structure of the vector bundle\rq{}s typical fiber  in a frame independent way; the gauge group is no longer postulated as fundamental: it acquires meaning as the invariance group of the typical fiber of $E$. So  we can think of parallel transport in a frame-independent way as being a structure preserving map, carrying the fiber's structure from one point of spacetime to another along a spacetime path. This was illustrated in the case of spinors, which are intimately tied to the Lorentzian structure of $TM$ and thus, consequently, to the structure group $SO(3,1)$. But it is the spinor\rq{}s geometric structure that determines its parallel transport, independently of a characterisation in terms of the bundle of frames---convenient though that characterisation might be. 

So, in Section \ref{sec:parallel_fiber} I will elaborate how parallel transport, even when explicitly articulated in terms of the frame bundle, can be abstractly characterised as applying to the entire fibers, in a frame-independent manner. In Section \ref{sec:Ehresmann} I will describe how we can, from an Ehresmann connection $\omega$ on $P$, obtain an abstract affine connection for vector bundles over $M$, in a frame, or trivialisation-independent, way. 

\subsubsection{Parallel transporting the entire fiber}\label{sec:parallel_fiber}
Structure groups play an important role in determining  the affine connection. This is clear from what we established earlier:  the structure group of a frame bundle is determined as that which preserves the structure of the typical fiber of the corresponding vector bundle. Thus, in general relativity, the reason we obtain an $SO(3,1)$ action on the space of frames is that each fiber $T_xM$ and its tensor products has a Lorentzian inner product structure. Talk about $SO(3,1)$ invariance of a certain vector bundle is  talk about Lorentzian metrics in another language. The role of $SO(1, 3)$ becomes \emph{explicit} only when we try to compute things explicitly, i.e. in a frame or coordinate basis. \emph{Mutatis mutandis}, the same applies for Yang-Mills theory. 

As to the connections, in both general relativity and Yang-Mills theory,  the space of connection-forms is isomorphic to the space of covariant derivatives---the difference between two covariant derivatives gives a connection-form; a particular isomorphism is fixed once we choose an origin, e.g. $\nabla_o$. Thus, in both general relativity and Yang-Mills, there is no difference between taking either a covariant derivative  or  a connection-form  as the structure that determines parallelism.

We wean  parallel transport from frame dependence by taking it as a structure-preserving transformation of the entire internal space, as we move from one spacetime point to another. In an abstract sense, we do not need to specify a basis because we do not need to think of parallel transport as applying to  single elements of the fiber: it is the entire fiber that is transported; and that applies to the $\bb C^2$ space of spinors (because it applies to $TM$) as well as to the $\bb C^3$ space of (one flavor of) quarks.  

 In more detail: in the principal bundle formalism of Section \ref{sec:parallel}, I have described parallel transport along a curve  $\gamma: [0,1]\rightarrow M$ via a horizontal lift $\gamma^h$ of $\gamma$ through a particular initial point, or frame, $\gamma^h(0)=p\in \pi^{-1}(\gamma(0))$. So one might be inclined to think that parallel transport in the principal fiber bundle language requires the stipulation of an initial frame, $p\in \pi^{-1}(\gamma(0))$. This would make the notion very much dependent on the use of frames.
 
  But the horizontal lift commutes with the group action,  $\gamma^h\circ L_g=L_g\circ \gamma^h$ (which follows from horizontal curves being sent to horizontal curves by translation of the origin; cf. \cite[Ch. II Prop. 3.2]{kobayashivol1}).  That means we can think of parallel transport without frames, as an isomorphism of an initial to a final fiber, e.g. for the path $\gamma$: 
  \be\label{eq:tau} \tau_\gamma: \pi^{-1}(\gamma(0))\rightarrow \pi^{-1}(\gamma(1)).
  \ee
   If we take $P$ as a frame bundle of a vector bundle $(E, \pi_E, F, M)$, we can equivalently write \eqref{eq:tau} as: 
     \be \tau_\gamma: \pi_E^{-1}(\gamma(0))\rightarrow \pi_E^{-1}(\gamma(1)).
  \ee
   
   The conclusion is that, both for spacetime and fiber bundles,  covariant derivatives can be characterized invariantly, without mentioning frames, gauges, etc. In the same way we think of the Levi-Civita connection as determining the rotation of the local tangent space as one moves from one point to another (and not as the explicit transport of a specific tangent vector), we think of the Ehresmann connection as determining the rotation of internal  spaces. So how is the Ehresmann connection represented on spacetime?

\subsubsection{The Ehresmann connection on spacetime: the Atiyah-Lie connection.}\label{sec:Ehresmann}
Here we want to show explicitly how the Ehresmann connection projects down to spacetime, without talking about trivializations (and frames). The only outstanding difference, it seems, between a Levi-Civita and an Ehresmann connection, is that the latter  lives in a principal bundle that has no unique representation at each spacetime point. This is only a superficial difference, as I will now show in detail.

As I described above, parallel transport is determined by horizontal directions in the bundle and  we know that the horizontal bundle $H\subset TP$, is left-invariant.  So, if we know what parallel transport is at $p$, we know what it is at $g\cdot p$. By getting rid of this redundancy, we can find a global spacetime representation of the connection $\omega$. 
To do that, we first note that there is a 1-1 relation between (Ehresmann) connection-forms and \textit{left-invariant} sections of $TP$ (see \cite[Ch. 4]{kobayashivol1}). 

Left-invariant vector fields are not unconstrained sections of the vector bundle $TP$, i.e. $C^\infty(TP)$. But they \emph{are} unconstrained sections  of $TP/G$, the so-called \textit{bundle of connections} (see e.g. \cite[Sec. 3.2]{Ciambelli}; \cite[p.9]{LeonZajac}; \cite[p.60]{sardanashvily2009fibre}; \cite[Ch. 17.4]{Kolar_book} and \citep{Jacobs_PFB}).\footnote{The bundle of connections appeared almost simultaneously in  \cite{AtiyahLie} and  \cite{Kobayaschi_bundle}. It is often  referred to as the \emph{Atiyah-Lie bundle}. See also \cite[Ch. 17.4]{Kolar_book}. To avoid confusion, it is better to refer to a section of the bundle of connections, which is itself a generalization  of a connection to what are known as Lie algebroids (see \cite{mackenzie_2005}), as an Atiyah-Lie connection.}  In other words, the difference between sections of $TP$ and $TP/G$  is that, while both can be seen as sections over $TM$ (with $\pi_*$ the projection), the latter---$TP/G$---is more constrained, since it can only encode left-equivariant objects defined on the first, $TP$.

The main idea in the construction of this bundle  is to take the projection map $\pi_*:TP\rightarrow TM$, and make it `forget' at which point or ``height'' of the orbit   it was applied.  The formalism  represents parallel transport of internal quantities for the directions in spacetime, rather than for directions in the bundle $P$.  
Thus $TP/G$ is most naturally a vector bundle over $TM$ rather than over $M$ or  $P$. But since $TM$ is itself a bundle over $M$, $TP/G$ can  also be construed as a bundle over $M$.

To define the fiber of $TP/G$, recall that a point in $TP$ is
locally described by $(p, v_p)$ with $v_p \in T_pP$, and the group $G$ acts (freely and transitively)  as
$(p, v_p)\mapsto (g\cdot p, {L_g}_*(v_p))$, which is the relation by which we define the left-invariant vector fields. Thus $TP/G$ is defined in much the same way as the associated bundle is defined from a vector space and a principal bundle, i.e. by identifying
\be(p, v_p) \sim (g\cdot p, {L_g}_*(v_p)), \quad \text{for all}\quad g\in G.\ee
Since locally (i.e. given some trivialization of the tangent  bundle) for $x=\pi(p)$ and $\xi\in \mathfrak{g}$, we can represent $p=(x, g):=g\cdot \sigma(x)$ and $v_p=(X_{x}, \xi):=\xi+\sigma_*(X_x)$ we have, locally, $(p, v_p)=(x,g, X_x, \xi)$. 
If we take the quotient, we obtain that the elements of the new vector bundle will be locally of the form $(x, X_x, \xi)$, as was to be expected from a Lie-algebra valued 1-form (or vector field).  


Given a point on $M$, and a tangent direction on $M$, and a local trivialization of the bundle, an element of the vector bundle $T^*P/G$  spits out a Lie-algebra element. Thus, as in the standard manner of obtaining ${A}^\sigma$ from $\omega$, here we also locally recover, in a trivialization, that the representative of the connection, call it $\Gamma$,  is the $\mathfrak{g}$-valued 1-form on $M$; 
$\Gamma$ is global, but in a local trivialization, it would be represented by $A_\mu^i$, where, the indices refer to a Lie-algebra and a tangent bundle basis.  So $\Gamma$ is like an abstract tensor, $X$, which stands to  its coordinate formulation  $A_\mu^i$, as $X$ stands to $X_{\mu_1\cdots\mu_n}^{\nu_1\cdots\nu_k}$.
The values of $\Gamma$ according to different trivializations are  related by the transformation \eqref{eq:gauge_transf}, just as the values of $X_{\mu_1\cdots\mu_n}^{\nu_1\cdots\nu_k}$ are related by coordinate transformations. Thus, the sections of the bundle $T^*P/G$ will be sections of a bundle over $M$, and they will be frame-invariant, and therefore, invariant under passive gauge transformations.

The advantage of a section of $T^*P/G$ over  the standard gauge potential is that it is globally defined and it is independent of internal coordinates (coordinates for the Lie algebra, and tangent bundle); and the advantage over the connection-form is that it is a section of a vector bundle with $T^*M$ as its base space. The disadvantage is that it is highly abstract. Nonetheless, this formulation allows a strong analogy between the basic kinematical variables of the gauge theory and the metric connection, in a coordinate-independent manner.  
 
 We can think of $\Gamma$, the section of the vector bundle $T^*P/G$, as one more physical field on spacetime. Since it is a section of a certain vector bundle, upon introducing coordinates (or frames) it admits changes of bases with which it is described, and these can be construed as  gauge transformations. But just as the connection $\omega$ is invariant with respect to these gauge transformations, so will be $\Gamma$.

\section{How to dispel the second disanalogy: the internal spaces}\label{sec:problems}\label{sec:dispel_second}

This Section will dispel the second putative disanalogy between parallel transport of spacetime and internal quantities. But before diving in, I want to clear the ground. 

The label `geometrical' might be taken to connote  properties related to distance relations, and to geodesics extremizing such distances. That is not how I mean it. Although there is one interpretation of gauge theories and gauge transformations that is geometric in this sense---called Kaluza-Klein theory (cf. \citep{KaluzaKlein} and \citep{Dawning_book} for the history)---that is not the sense I will focus on here. Here I want to assess whether gauge transformations can be understood naturally as automorphisms of an internal geometric structure; and whether  the Ehresmann connection can be understood similarly to an affine connection determining parallelism for tensor fields over spacetime.

 Next, let us set aside all questions about the `external\rq{} spacetime geometry. A matter field can be described as the tensor product of some inner space on which gauge fields take values, e.g. a  complex scalar or Yang-Mills field, $\phi$,
 and a spinor field $\psi$; or tensor fields in the case of gauge bosons.  
 The gauge fields use a connection and gauge frame which are independent of the spacetime manifold frame, while spinor and tensor fields mirror the connection and changes in frame of the spacetime manifold, as we saw in Section \ref{sec:spinors}. So gauge fields are acted on by representations of the gauge group and its Lie algebra, while spinor fields are acted on by representations of the Spin group and its Lie algebra ($\mathfrak{so}(3,1)$).  The gauge component then responds to gauge transformations, while the spinor component responds to changes of frame. Here, I will focus only on the gauge part. 

More specifically, in order to interpret the Ehresmann connection and gauge transformations as on a par with  the Levi-Civita connection we need to respond to Weatherall's second disanalogy: in the SM, different fields live in different spaces, and the Ehresmann connection lives outside of these spaces but plays an auxiliary role. We will see that interacting fields can be seen as  sections of bundles built up from the same internal spaces, or typical fibers. For instance, in the same way that a covariant tensor of rank two is built from two copies of $TM$, (the internal part of) quarks will have components in three copies of $\bb C^3$.  Thus, by describing the connection form $\omega$ in the bundle of admissible frames  of  $(E, M, \bb C^3)$, we have a geometric reason for the parallel transport of the different quarks and leptons marching in step. This will of course require a brief description of the particle content of the SM.  

In Section \ref{sec:intro_SM} I will introduce the  main ideas in a simplified model, containing only quarks and gluons. This illustrates how gluons can be accommodated once we have fixed the underpinning geometry of the typical fibers of the matter fields (the fermions), as described in Section \ref{sec:parallel}.  In Section \ref{sec:SM}, I will include the entire content of the SM. 

\subsection{Inhabiting the same space}\label{sec:intro_SM}

 Let us here focus solely on the QCD Lagrangian. In this first Sectiojn I will only sketch the argument, giving more details in the next Section. In the QCD Lagrangian, the $SU(3)$ connection on a local trivialisation is called $G^i_\mu$, with $i$ indices in $\mathfrak{g}=\mathfrak{su}(3)$, and it shows up twice: once via minimal coupling to the quarks, and once in the self-interaction terms. The quarks are Dirac spinors, $\psi_\alpha$, with indices representing colour; and it is standard to have $\mathfrak{su}(3)$ act on them via its generating Gell-Mann 3x3 matrices, $t^i$: these are the matrices representing $\mathfrak{su}(3)$ on $\bb C^3$ (the so-called \emph{fundamental} representation). So we have in the Lagrangian terms of the form (one for each quark type): 
\be L_{\text{\tiny QCD}}= \bar \psi^\alpha \gamma^\mu(\delta_{\alpha}^\beta\partial_\mu+  (G^i_\mu t_i)_\alpha^\beta) \psi_\beta + G^i_{\mu\nu}G_i^{\mu\nu} +m\bar \psi^\alpha \psi_\alpha,
\ee
where $t^i_{\alpha\beta}$ are the $\alpha\beta$ components of the Gell-Mann $i$-th matrix. 

Crucially, though in the kinetic term $G^i_{\mu\nu}$ `lives' in  a different vector bundle than $\psi^\alpha $, namely, the bundle with $\mathfrak{g}$ as the typical fiber, we could still write down the same kinetic term as: 
\be (G^i_{\mu\nu}t_i)_{\alpha\beta}(G_j^{\mu\nu}t_j)_{\alpha\beta},
\ee
because it is a property of these matrices that 
\be  t^i_{\alpha\beta}t^j_{\alpha\beta}=\delta^{ij}.
\ee

So, though it might be more cumbersome to use the same representation for all of the appearances of the connection-form, we could do that, in principle. Of course, in this simplified example  I only really considered one kind of quark (or rather, I assumed all of the quarks were constructed from the same space, which here I left unspecified), so our success was guaranteed by the interpretation of connection forms $\omega$ in terms of frame bundles for the quarks, as described in Section \ref{sec:PFB_formalism}. In the next Section, we will assume that, once we figure out the typical fiber of the matter fields,  the connection-forms can be effortlessly accommodated in a similar way. 


\subsection{The typical fibers of the SM}\label{sec:SM}
As illustrated in the previous Section, and explained in Section \ref{sec:PFB_formalism} (see also Section \ref{sec:parallelism}, we don\rq{}t need to here worry about the representation of the connection-forms in the SM: 
 their meaning is derivative from that of the typical fibers. In particle physics jargon, connections are the `force-carriers\rq{}, and are represented by gauge bosons; so here, I will put the bosons aside and  focus on the fermionic content of the SM.

 In the SM as a whole, there are 48 Weyl fermions, which, as we saw in Section \ref{sec:spinors}, are two-component spinors, or elements of $\bb C^2$.  But I am only interested in the structure of the internal spaces; the spaces where the gauge connections act. So here I am basically ignoring the spacetime spinor structure of the SM (though they are somewhat implicit in the notation of left or right handed particles to be used below). When representing the full fermionic content of the SM, this spinor part would be included as factors in a tensor product with the internal part that I am interested in and aim to describe in this Section.\footnote{What I am saying here would be strictly true if I represented the SM solely in terms of one chirality, which is certainly possible. So, instead of having both right and left handed spinors, one can include in the table only left-handed ones. This would have the advantage of being rigorous about the tensor product between internal spaces and spinors but would have the disadvantage of having to introduce complex conjugates of the representations, e.g. use $\bar 3$ instead of $3$ for the  first and fourth row of the table below, and change other representations all around. But this is a rather trivial technical point, and thus I have chosen to stick to the simpler fundamental representation and use both right and left-handed spinors.   } 
 
 The part of the SM that I am interested in consists of 48 complex numbers, organised into three generations, which means it has the same structure repeated three times. We can understand this repetition in terms of direct sums:
\be \bb C^{48}=\bb C^{16}\oplus \bb C^{16} \oplus \bb C^{16}\ee
The table below tells us how these components transform, and it is organised into blocks of elements that can transform into each other (which the three different generations cannot). So each $\bb C^{16}$ breaks down into the six rows of the table:\footnote{Here I am including the right-handed neutrinos, which have not yet been directly observed, but, after the discovery of neutrino oscillations, are generally assumed to exist. The \emph{minimal} SM would not have the last row. }

\begin{center}\begin{tabular}{c|c|c|c|}
  ~&  $SU(3)$ & $SU(2)$ & $U(1)$ \\
  \hline
 $q_L$& 3 & 2 & $\frac{1}{6}$ \\
  \hline
  $u_R$& 3 & 1 & $\frac23$ \\
  \hline
   $d_R$& 3 & 1 & $-\frac13$ \\
  \hline
   $\ell_L$& 1 & 2& $-\frac12$ \\
  \hline
   $e_R$& 1 & 1 & 1 \\
  \hline
   $\nu_R$& 1 & 1 & 0 \\
  \hline
\end{tabular}
\end{center}

Now let us unpack it: 
\begin{itemize}
\item \textbf{The quarks:} are represented by the first three rows of the table. $q_L$ is a left-handed quark doublet, which is a doublet of the form $q_L=(u_L, d_L)$. In the first generation this would be called up-left and down-left, respectively; in the second generation it would be charm-left and strange-left, and in the third generation it would top-left and bottom left. The reason $q_L$ is called a doublet is that the components of $q_L$, namely $u_L$ and $d_L$, are charged under the weak nuclear force, and transform into each other under the action of $SU(2)$; unlike the two rows beneath it, representing the up-right and the down-right quarks, $u_R$ and $d_R$ which are singlets. In the entry corresponding to $q_L\times SU(2)$ this transformation property is represented by the number 2, which  means that $q_L$ transforms as an element of $\bb C^2$ under the fundamental representation of $SU(2)$. The number 1 for the entries $u_R\times SU(2)$ and $d_R\times SU(2)$ means that $u_R$ and $d_R$ are neutral under the weak forces, so cannot transform into each other (because, being singlets,  they don\rq{}t transform at all under $SU(2)$). And as to the first column: quarks clearly feel the strong forces, and they transform as elements of $\bb C^3$ under the standard, or fundamental, representation of $SU(3)$. Finally, the left-handed quark has a `hypercharge\rq{} of $-1/6$ under $U(1)$, which means that it is a complex number (an element of $\bb C$) which under the action of a given $U(1)$ phase shift generator $\xi$, has its phase rotate at the rate of $-\xi/6$ (or $e^{i\xi/6}$); \emph{mutatis mutandis} for the down-right and up-right quarks.\footnote{Note that for $U(1)$ it is a 0 entry---and not a 1, as it is for $SU(3)$ and $SU(2)$---that tells us a particle does not transform, or is neutral with respect to this interaction.  I should also note that hypercharge is not exactly the same as electric charge. It coincides with electric charge only for the rows that transform trivially under $SU(2)$, namely, for all the right-handed particles in the table. Thus the electric charge of the down-right quark is $-1/3$, for an up-right quark it is $2/3$, etc. These particles don\rq{}t feel the weak interactions; for those that do, namely $q_L$ and $\ell_L$, the electric charge emerges only after symmetry-breaking, and it is a combination of weak and hypercharge. The way these charges combine after symmetry-breaking gives a mnemonic device for the numbers of the last column: the entry for the left-handed particles is the average of the two entries below, for right-handed particles.\label{ftnt:hyper}}  
\item \textbf{The leptons:} are represented by the remaining three rows in the table and have a kind of parallel structure to the quarks, but of course they are all neutral under $SU(3)$ (they are not charged under strong interactions).  $\ell_L$ is the left-handed lepton doublet, which is of the form $\ell_L=(e_L, \nu_L)$. In the first generation these are the left-handed electron and neutrino (in the second and third they get appended by muons and taus). Again, we put $e_L$ and  $\nu_L$ in the same row because they are charged under $SU(2)$ (they are charged under the weak forces), and  transform into each other, unlike the particles of the two remaining rows---the right-handed electron and neutrino, $e_R$ and $\nu_R$, which are neutral under $SU(2)$.  The hypercharge of $\ell_L$ is $-1/2$ (which does not coincide with its electric charge; see footnote \ref{ftnt:hyper}). The electric charge of the right-handed electron (or positron), is, as expected, 1, and the right-handed neutrino has no electric charge. 
\end{itemize}

Thus the first row is represented as an element of $\bb C^3\otimes \bb C^2\otimes \bb C$: the left handed quark doublet  has components lying along these internal spaces: we must locate it within a space of three colours, and of two isospin charges, and of one hypercharge. Unlike the first row, the particle in the following two rows have no component along $\bb C^2$, which is why it is not charged under $SU(2)$. So e.g. the down-quark has three options for colour, and  only one option for isospin and electric charge. In contrast, the left-handed lepton doublet has no components along $\bb C^3$, but has components along $\bb C^2$; and the left-handed electron and neutrino have no components along either $\bb C^3$ or $\bb C^2$: that is why they are not charged under either the strong or the weak interactions, etc. 
So can conceive of each generation\rq{}s $\bb C^{16}$ as having the following decomposition into six (non-identical) factors: 
\begin{eqnarray}
 \bb C^{16}&=&(\bb C^3\otimes \bb C^2\otimes \bb C^1)\oplus (\bb C^3\otimes \bb C^1\otimes \bb C^1)\oplus (\bb C^3\otimes \bb C^1\otimes \bb C^1)\nonumber\\
 &~&\oplus (\bb C^1\otimes \bb C^2\otimes \bb C^1)\oplus (\bb C^1\otimes \bb C^1\otimes \bb C^1)\oplus (\bb C^1\otimes \bb C^1\otimes \bb C^1).
\end{eqnarray}

And we can finally answer the main question of this Section, and indeed of the paper: why do the parallel transports of different particles, as sections of different vector bundles, march in step? Because, just as tensor bundles are constructed from the underpinning geometry of $TM$ (possibly endowed with an inner product and orientation in more complicated cases, like that of the spinor), and tensors have components in the spaces thus constructed, particle fields have components in internal spaces corresponding to colour, isospin, and (hyper)charge, that are constructed from the underpinning geometry of $\bb C^3\otimes \bb C^2\otimes \bb C^1$, endowed with an inner product and, except in the case of $\bb C^1$, an orientation. The structure groups $SU(3)\times SU(2)\times U(1)$ are the symmetries that preserve the structure of these internal spaces, or typical fibers.  

Abstractly, the affine connections---in particle physics terminology called the gluon, the W and the Z-bosons---dictating the parallel transport of colour, isospin, and (hyper)charge along a certain spacetime curve $\gamma :[0,1]\rightarrow M$ take   $\bb C^3\otimes \bb C^2\otimes \bb C^1$ over $\gamma(0)\in M$ to $\bb C^3\otimes \bb C^2\otimes \bb C^1$ over $\gamma(1)\in M$, as a linear, structure preserving transformation. As with spinors, one can make this parallel trasport explicit by choosing frames for these spaces, in which case we obtain a gauge redundancy corresponding to different choices of frames. Namely, we construct the bundle of frames over the vector bundles $(E, M, \bb C^3), (E, M, \bb C^2), (E, M, \bb C^1)$, and have  connection-forms $\omega_g, \omega_w, \omega_z$ dictate the parallel transport of those frames. But, at an abstract level, we could characterise everything in geometric terms, without coordinates or frames.

Let me address a possible overall objection: given the Lagrangian of the standard model in a certain coordinate system, I could extract all of the invariances and symmetry transformations directly. The Lagrangian would constrain the internal values of the different particle fields to appropriately co-rotate. This is a true statement, but I don\rq{}t think it is explanatory. For the same could of course be said about general covariance in general relativity. There, it is the geometric interpretation that underpins the universal coupling of all of the fields to the spacetime geometry. But this universality could fail; for instance, if   `bi-metric\rq{} theories of gravity were adopted. The fact that such bi-metric theories have little empirical support can be explained by the  more parsimonious, familiar geometric interpretation of general relativity. Similarly, my argument here shows  that the most parsimonious explanation for the current form of the standard model (without the analogous `bi-metrics\rq{}), is that it concerns an internal structured space, $\bb C^3\otimes \bb C^2\otimes \bb C^1$. 

\section{Conclusions and outlook}\label{sec:conclusions}

In particle physics,  \emph{fundamental forces}  are uniquely associated with structure groups. I have argued here that those structure groups merely reflect the structure of internal spaces in spacetime. The standard model  cleanly illustrates this idea. In just the same way that a Lorentzian inner product on  the tangent bundle $TM$ leads directly to the  local symmetry group $SO(3,1)$, the geometric structure of the internal spaces in which the fundamental particles take their values---$\bb C^3\otimes \bb C^2\otimes \bb C^1$ endowed with an inner product and orientation---leads directly to the familiar local symmetry group $SU(3)\times SU(2)\times U(1)$ representing the fundamental forces.  Any particle field that interacts with a fundamental force has components in the corresponding internal space; as we move from one point of spacetime to another, the standard of constancy for that internal space will dictate the parallel transport of those components. 

Thus I conclude that gauge theory is geometrical in a very strong sense, but its geometry is that of internal vector spaces, not of principal fiber bundles.  Agreed, in many applications  a description of a theory using  principal fiber bundles  may be more  practical (as it often is with spinors; cf.  Section \ref{sec:intro_spin}). Principal bundles have symmetry groups at their core, and  would successfully coordinate  symmetry transformations and parallel transport in their associated vector bundles; but they are redundant if the fields are known to inhabit internal, geometrically structured spaces. 

In light of this conclusion, one could replace \citet[p. 2401]{Weatherall2016_YMGR}\rq{}s metaphor, that \lq\lq{}Principal bundles are auxiliary [...] in the sense in which a coach is auxiliary to the players on the field\rq\rq{}\footnote{Because, as he says: \lq\lq{}vector bundles represent possible local states of matter; principal bundles   coordinate between these vector bundles\rq\rq{}.} with another metaphor, drawn not from sports but from music.  Just as all agree that, in their public performances, after arduous preparation,  a top-quality orchestra such as the Vienna or Berlin Philharmonic hardly needs the conductor, who is by then almost an epiphenomenon, so also in gauge theories, the vector bundles play all the music and the principal fibre bundle is almost an epiphenomenon.\footnote{There is a close analog here to   the debate between the dynamical and the geometric views on Lorentz-invariance (cf. \citep{Brown_book} for an extended defence of the dynamical approach, and \citep{BrownRead_dyn} for a recent survey). Roughly, that debate focuses on an order of explanation: are dynamical laws (locally) Lorentz invariant \emph{because} they at most survey a geometric landscape that is Lorentz-invariant, or is such a  geometry just a convenient way to codify Lorentz-invariant dynamical laws?  Transposing that debate to gauge theory:  since thus far it lacked a comprehensive geometric framework that was on  a par with its relativistic cousin, gauge theory might have been more favorable to the  dynamical view of symmetries. Now  gauge theory finds a natural home within (at least a very close analog of!) the geometric view. }  

Indeed, the new viewpoint achieved in this paper opens up a novel interpretative project for gauge theory as a whole. Major lines of enquiry, such as: empirical significance of gauge symmetries (cf. e.g. \citep{BradingBrown, GreavesWallace, Gomes_new}); the Higgs mechanism (cf. e.g. \citep{EarmanSSB2002, Smeenk2006,  Lyre2012}); non-locality and the Aharanov-Bohm effect (cf. e.g. \citep{healey1997ab, Maudlin_ontology, Wallace_deflating}),\footnote{This list is very far from exhausting! I merely tried to take samples from three different points of view on each issue. } etc; might profitably be reconstrued in terms of internal geometries.


\subsection*{Acknowledgements} I would like to thank Latham Boyle, for his patience and generosity in explaining to me just exactly what the typical fibers of the vector bundles in the standard model were. I would also like to thank: Aldo Riello, David Tong, and Tim Koslowski,  for a few conversations about this topic, and Jeremy Butterfield for comments and the suggestion of a new metaphor. 
\appendix
\section*{APPENDIX}

\section{The connection-form}\label{app:connection}

 In more detail: 
Given an element $\xi$ of the Lie-algebra $\mathfrak{g}$, and the action of $G$ on $P$, we use the exponential to find an action of $\mathfrak{g}$ on $P$. Taking the tangent of this map,  we define an embedding of the Lie algebra into the tangent space, given by the \emph{hash} operator: $\#_p: \mathfrak{g}\rightarrow T_pP$. So the vertical space $V_p$ at a point $p\in P$,  that is tangent to the orbits of the group, is also the image of this map; it is the linear span of vectors of the form 
\be\label{eq:fund_vec} \text{for}\quad \xi\in \mathfrak{g}:\quad {\xi^\#}(p):=\frac{d}{dt}{}_{|t=0}(\exp(t\xi)\cdot p)\in V_p\subset T_pP.\ee
Vector fields of the form $\xi^\#$ for $\xi\in \mathfrak{g}$ are called \emph{fundamental vector fields}.\footnote{It is important to note that there are vector fields that are vertical and yet are not fundamental, since they may depend on $x\in M$ (or on the orbit). Nonetheless, fundamental vector fields serve, together with the horizontal vector fields, as a basis for all the vector fields $\mathfrak{X}(P)$. \label{ftnt:vertical}} 

With these notions at hand, we define: 
\begin{defi}[An Ehresmann connection-form]  $\omega$ is defined as a Lie-algebra valued one form on $TP$, satisfying the following properties:
\be\label{eq:omega_defs}
\omega(\xi^\#)=\xi
\qquad\text{and}\qquad
{L_g}^*\omega=\Ad_g\omega,
\ee
  where the adjoint representation of $G$ on $\mathfrak{g}$ is defined as $\Ad_g\xi=g\xi g^{-1}$, for $\xi\in \mathfrak{g}$;  ${L_g}^*\omega_p(v)=\omega_{g\cdot p}({L_g}_* v)$ and where ${L_g}_*$ is the push-forward of $TP$ induced by the diffeomorphism  $g:P\rightarrow P$.\label{def:Ehres}\end{defi}

The vertical spaces are canonically defined by the action of the group; there is no choice there. On the other hand, the choice of connection is equivalent to a choice of covariant `horizontal' complements to the vertical spaces, i.e. $H_p\oplus V_p=T_pP$, with $H$ compatible with the group action, so that ${L_g}_*H_p=H_{g\cdot p}$. Since $\omega$ is $\mathfrak{g}$-valued and gives an isomorphism between  $V_p$ and $\mathfrak{g}$ ($\omega$'s  inverse is $\#: \mathfrak{g}\mapsto V\subset TP$), the first condition of \eqref{eq:omega_defs} means that we can define a horizontal space as the kernel $\mathsf{Ker}(\omega_p)=:H_p$. The second condition of \eqref{eq:omega_defs} guarantees that the notion of horizontality covaries with the choice of representative of the fiber (e.g. the choice of frame in the frame bundle example above), that is: a vector $v\in T_pP$ is horizontal iff ${L_g}_* v\in T_{g\cdot p}P$ is horizontal.\footnote{ That is, $\mathsf{Ker}(\omega_p)$ and  $V_p$ are transversal---$V_p$ has no element in $\mathsf{Ker}(\omega_p)$---and $\mathsf{Ker}(\omega_p))\oplus V_p=T_pP$. Since $\pi_*(T_pP)=T_{x}M$, and $\pi_*V_p=0$, we get $\pi_*(\mathsf{Ker}(\omega_p))=T_{x}M$, which ensures the horizontal spaces are isomorphic to the tangent spaces of $M$. Or, alternatively, since  \be\text{dim}(\mathsf{Ker}(\omega_p))+\text{dim}(\mathsf{Im}(\omega_p))=\text{dim}(T_pP)=\text{dim}(T_xM)+\text{dim}(\mathfrak{g}),\ee we obtain $\text{dim}(\mathsf{Ker}(\omega_p))=\text{dim}(T_xM)$.  Thus the vectors spanning $\mathsf{Ker}(\omega_p)$ are the so-called \textit{horizontal} vectors in the bundle, and each represents a unique `horizontal lift' at $p$ of a direction at $T_{\pi(p)}M$.  A horizontal lift of a vector $X_x\in T_xM$ through $p\in \pi^{-1}(x)\subset P$ is a horizontal vector $X^h_p$ such that $\pi_*X^H_p=X_x$. It is easy to show that (see \cite[Prop 1.3]{kobayashivol1}): (i) the lift of $X+Y$ is $X^h+Y^h$, (ii) $f^hX^h$ is the lift of $fX$, where $f^h:=f\circ\pi$, (iii) $\hat H(\lbr X^h, Y^h\rbr) $ is the horizontal lift of commutator of vector fields on $M$ $\lbr X ,Y\rbr$ (this is the only non-trivial item, but it is easy to prove: for  $\hat H(\lbr X^h, Y^h\rbr)$ is horizontal, and $\pi_*\hat H(\lbr X^h, Y^h\rbr)=\pi_*(\lbr X^h, Y^h\rbr)=\lbr X ,Y\rbr$ from the first two items. } 
  
  And, as usual, parallel transport may not commute along different horizontal directions, and the failure to commute is described by the \emph{curvature two-form}:
\be\label{eq:Omega}\Omega^i_j:=\d \omega^i_j +\omega^i_k\wedge\omega^k_j
\ee
If the curvature is non-zero, no choice of  frame over spacetime can be everywhere horizontal.

\begin{figure}[h!]\label{fig1}
\center
\includegraphics[width=0.8\textwidth]{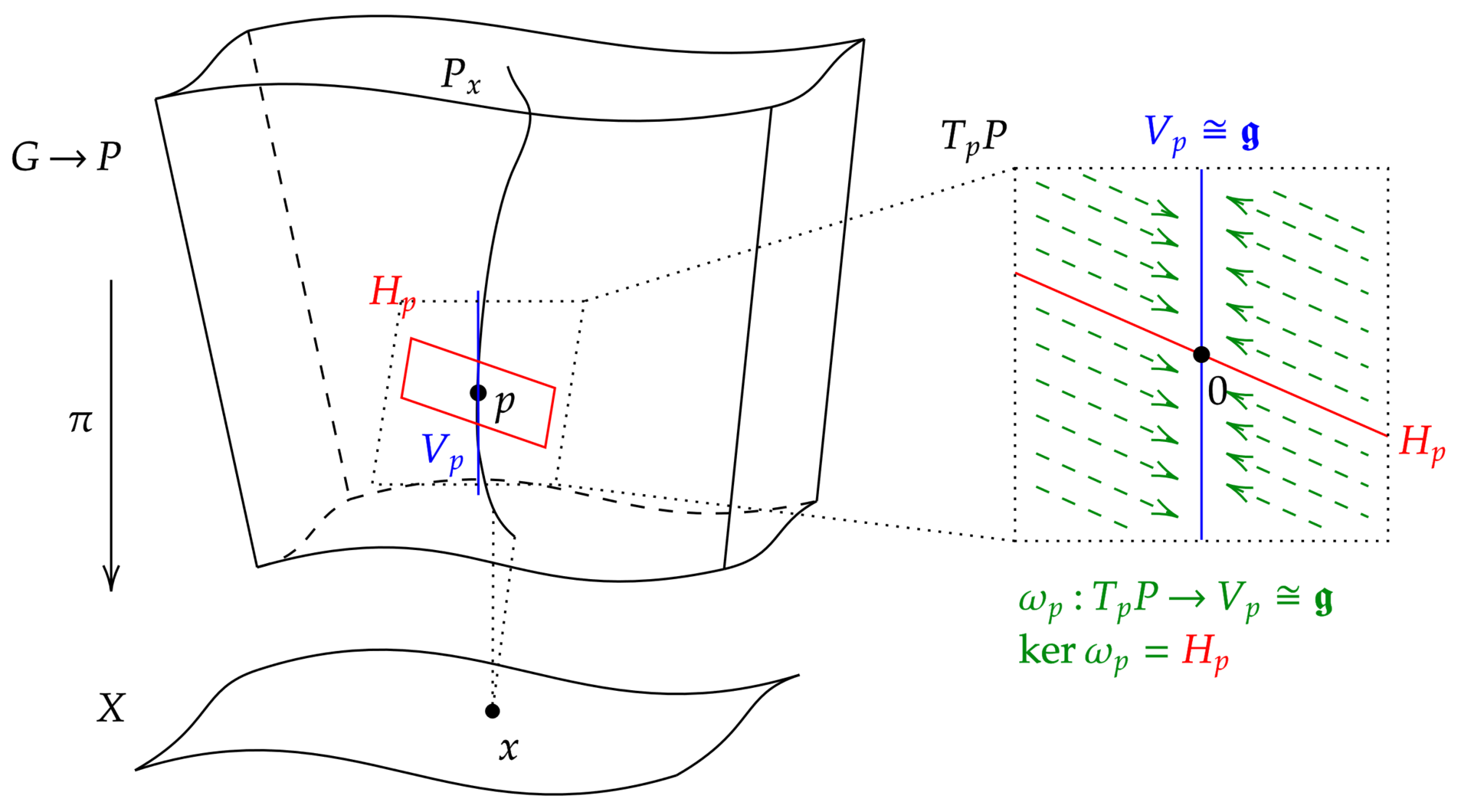}
\caption{ The relation between the Ehresmann connection form $\omega$ and a vertical projection, on the principal fiber bundle with structure group $G$.}
\end{figure}

To illustrate the encoding of parallel transport of tangent vectors as horizontal directions of a principal bundle, we  return to the tangent bundle $TM$, from the frame bundle, $L(TM)$. 
Once we have constructed associated bundles in this way, parallel transport for any vector bundle comes naturally from a notion of horizontality in the principal bundle. To find the parallel transport of the vector $X_x$ along $Y_x$, take the curve $\gamma(t)\in M$ with $\gamma(0)=x$, and so that $\gamma'(0)=Y_x$. Given a frame $p_x\in P$ so that $\pi(p_x)=x$, we take the horizontal lift of $\gamma(t)$ through $p_x$: call it $\tilde \gamma(t)$ (this is the unique curve that starts at $p_x$, projects down to $\gamma$, and whose tangent is everywhere horizontal, where uniqueness follows from the linear isomorphism $\pi_*:H_p\rightarrow T_{\pi(p)}M$: cf. \cite[Prop. 3.1]{kobayashivol1}). Let $X_x=[p_x, v]$, where $v\in \RR^n$ are the components of $X_x$ in terms of the basis $p_x$. 
By definition, the curve in $E$ given by $[\tilde\gamma(t), v]$ is parallel transported, i.e. gives a parallel transport of $X_x$ along $\gamma(t)$. Now, we can define the covariant derivative of a vector field  $X$  such that $X(x)=X_x$ as follows. First, we define $v_X:P\rightarrow \RR^4$   such that, for all $p\in P$
\be\label{eq:vectorP} X(\pi(p))=[p, v_X(p)],\quad \text{where}\quad v_X(g\cdot p)=g^{-1}v_X(p);\ee
that is,  $v_X(p)$ are the components of $X(\pi(p))$ on the basis $p$ (and  therefore $v_X$ obeys the covariance property on the right of \eqref{eq:vectorP}). Thus we define the covariant derivative of $X$ along $Y$ at $x$, as:
\be \D_YX(x):=\lim_{t\rightarrow 0}\frac{1}{t}([\tilde\gamma(-t), v_X(\tilde\gamma(-t))]-[p_x, v_X(p_x)]).
\ee
In words, we compare the parallel transported components of $X$ with the actual components of $X$; their non-constancy corresponds to the failure of $X$ to be parallel transported, and to the non-vanishing covariant derivative of $X$.  In this way a covariant derivative is just the standard derivative of the vector components as described in the horizontal---or parallel transported---frame. This is, in words, the description of the covariant derivative of $X$ along $Y$ at $x\in M$.

\bibliographystyle{apacite} 
\bibliography{references3}

\end{document}